\documentclass[journal,twoside]{IEEEtran}

\newcommand{\comment}[1]  {  }
\def\BE{\begin{equation}}
\def\EE{\end{equation}}
\def\BEA{\begin{eqnarray}}
\def\EEA{\end{eqnarray}}
\newcommand{\pd}[2]{\frac{\partial #1}{\partial #2}}

\newcommand\vx{{\bf x}}

\usepackage{epsfig}
\usepackage{graphicx}  
\usepackage{amssymb}
\usepackage{amsmath}    
\begin{document}
%
\title{Capacity of Complexity-Constrained\\Noise-Free CDMA}
%
%
\author{Ori~Shental,~\IEEEmembership{Student Member,~IEEE,}
        Ido~Kanter,
        and~Anthony~J.~Weiss,~\IEEEmembership{Fellow,~IEEE}
\thanks{Manuscript received July 12, 2005. The associate editor coordinating the
review of this letter and approving it for publication was Prof. Giorgio Taricco.
        This research was supported by the Israel Science Foundation (Grants 1232/04, 296/03).}
\thanks{O. Shental and A. J. Weiss are with the Department of Electrical Engineering-Systems,
Tel-Aviv University, Tel-Aviv 69978, Israel (e-mail: \{shentalo,ajw\}@eng.tau.ac.il).}
\thanks{I. Kanter is with the Department of Physics, Bar-Ilan University, Ramat-Gan 52900, Israel (e-mail:
kanter@mail.biu.ac.il).}}

%
%
%
\markboth{IEEE COMMUNICATIONS LETTERS, VOL. x, NO. y, MONTH 200z}{Shental \MakeLowercase{\textit{et al.}}:
Capacity of Complexity-Constrained Noise-Free CDMA}
%



\maketitle

\begin{abstract}
An interference-limited noise-free CDMA downlink channel operating under a complexity constraint on the receiver
is introduced. According to this paradigm, detected bits, obtained by performing hard decisions directly on the
channel's matched filter output, must be the same as the transmitted binary inputs. This channel setting,
allowing the use of the simplest receiver scheme, seems to be worthless, making reliable communication at any
rate impossible. We prove, by adopting statistical mechanics notion, that in the large-system limit such a
complexity-constrained CDMA channel gives rise to a non-trivial Shannon-theoretic capacity, rigorously analyzed
and corroborated using finite-size channel simulations.
\end{abstract}


%
\IEEEpeerreviewmaketitle

\section{Introduction}
%
%
%
%
\PARstart{D}{irect-sequence} spread-spectrum code-division multiple-access (CDMA) is used extensively in modern
wireless communication systems and serves preeminently in commercial cellular networks. Investigation of
reliable (\textsl{i.e.} errorless) communication via the CDMA channel is a long-standing and productive research
topic (\textsl{e.g.},~\cite{BibDB:VerduShamai}).

A typical investigation of a CDMA channel often assumes an upper bounded transmission power, but no restrictions
on complexity are imposed. In the era of ubiquitous and pervasive communications, in, for example, indoor and
personal area networks (PAN), there is an emerging interest in a complementary scenario. According to this
scenario, the CDMA system operates in a high signal-to-noise ratio (SNR) regime (thus power limitation is less
crucial), but is highly restricted by its receiver's signal processing complexity. For instance, this is the
case in complexity-limited (rather than noise-limited) CDMA downlink, where the simplest mobile receiver is to
be used. Such a trivial receiver requires that detected bits, sliced at the output of the channel's matched
filter, must be the same as the transmitted binary inputs. Formerly, there has been no examination of the
information-theoretic characteristics of CDMA channels in this setting, being especially applicable for the
downlink.

In this contribution, we compute the Shannon capacity of such a complexity-constrained CDMA channel with binary
signaling, random spreading and arbitrary user load. For this purpose, we borrow analysis tools from equilibrium
statistical mechanics, especially the Hopfield model of neural networks~\cite{BibDB:Gardner,BibDB:Singh}.
The achievable asymptotic information rates of such naive CDMA channels, which may seem worthless from an
information-theoretic point of view, are found to result in valuable rates, comparable to those achieved by
using the optimal multi-user receiver.

\section{Channel Model}\label{sec_model}
Consider a noiseless synchronous CDMA downlink accessing $K$ active users via the mutual channel in order to
transmit their designated (coded) information binary symbols, $x_{k}=\pm1$, where $k=1,\ldots,K$. Each
transmission to a user is assigned with a binary signature sequence (spreading code) of $N$ chips, $s_{k=1\ldots
K}^{\mu=1\ldots N}=\pm1$. Assuming a random spreading model, the binary chips are independently equiprobably
chosen, and the deterministic chip waveform has unit energy. The cross-correlation between users' transmissions
is $\rho_{ki}\triangleq1/N\sum_{\mu}s_{k}^{\mu}s_{i}^{\mu}$. The received signal is passed through the user's
matched filter. Thus, the overall channel input-output relation is described by \BE\label{eq_channel}
    y_{k}=x_{k}+\sum_{i\neq k}\rho_{ki}x_{i},
\EE where the $k$'th user matched filter output, $y_{k}$, is the designated bit, $x_{k}$, corrupted by an
interference term. This interference term is composed of a summation over (cross-correlation) scaled versions of
all other users' bits. The set of all cross-correlations $\rho_{ki}$ is hereinafter denoted by $\rho$. In the
following asymptotic analysis, we assume that $K\rightarrow\infty$, yet the system load factor $\beta\triangleq
K/N\triangleq\alpha^{-1}$ is kept constant, and that the information rate is the same for all users,
\textsl{i.e.} $R_{k}=R$.

We want to convey information reliably through the channel~(\ref{eq_channel}) under a low-complexity constraint
on the user receiver. According to this constraint, detected bits, $\hat{x}_{k}$, obtained by performing hard
decisions directly on the channel's matched filter output samples, must be the same as the transmitted bits.
Explicitly, $x_{k}\equiv\hat{x}_{k}=\emph{{\textrm{Sign}}}(y_{k})$, where $\textrm{Sign}(\cdot)$ is the trivial
sign function. Under the constraints outlined above it is clear that not all combinations of input symbols will
result in errorless communication. Thus, the capacity of the channel can be obtained by evaluating the number of
codewords that ensure errorless detection. Following, we prove that this complexity-constrained CDMA channel
setting yields non-trivial capacity.

\section{Capacity}\label{sec_capacity}
A binary codeword $\vx^{c}\triangleq\{x_{1}^{c},\ldots,x_{K}^{c}\}$, composed of all $K$ users' bits at a given
channel use, for which the channel constraints hold, satisfies the condition \BE\label{eq_stable}
    \int_{0}^{\infty}\prod_{k=1}^{K}d\lambda_{k}\delta(y_{k}-\lambda_{k}x_{k}^{c})=
    \int_{0}^{\infty}\prod_{k=1}^{K}d\lambda_{k}\delta(\alpha y_{k}-\lambda_{k}x_{k}^{c})=1\nonumber,
\EE where $\delta(\cdot)$ is the Dirac delta function. Let the random variable $\mathbb{N}(K,\beta,\rho)$ denote
the number of codewords, \textsl{i.e.} \BE\label{eq_number0}
    \mathbb{N}(K,\beta,\rho)\triangleq\int_{-\alpha}^{\infty}\prod_{k}d\lambda_{k}
    \sum_{\vx}\prod_{k}\delta\Big(\sum_{i\neq k}\alpha\rho_{ki}x_{i}-\lambda_{k}x_{k}\Big)\nonumber,
\EE where $\sum_{\vx}$ corresponds to a sum over all the possible values of the transmitted input symbols.
Assuming equal user information rates, the corresponding asymptotic capacity of the channel is
defined~\cite{BibDB:BookCover}, in bit information units, as
$C_{\infty}(\beta)\triangleq\lim_{K\rightarrow\infty}\log_{2}{\mathbb{N}(K,\beta,\rho)}/K$. According to the
self-averaging property~\cite{BibDB:BookEllis}, in the large-system limit, $K\rightarrow\infty$, the number of
successful codewords $\mathbb{N}(K,\beta,\rho)$ is equal to its expectation with respect to (w.r.t.) the
distribution of $\rho$, \textsl{i.e.} \BEA\label{eq_number1}
    \lim_{K\rightarrow\infty}\mathbb{N}(K,\beta,\rho)&=&\mathcal{N}(\beta)=
    \lim_{K\rightarrow\infty}\int_{-\alpha}^{\infty}\prod_{k}d\lambda_{k}
    \\&\times&\sum_{\vx}\Bigg<\prod_{k}\delta\Big(\sum_{i\neq
    k}\alpha\rho_{ki}x_{i}-\lambda_{k}x_{k}\Big)\Bigg>_{\rho},\nonumber
\EEA where $\mathcal{N}(\beta)$ and $<\cdot>_{\rho}$ denote the average and averaging operation, respectively.
Representing the delta function by the inverse Fourier transform of an exponent and substituting
$x_{k}\omega_{k}$ for the angular frequency of the Fourier transform $\omega_{k}$, expression~(\ref{eq_number1})
can be rewritten as \BEA\label{eq_number3}
    \mathcal{N}(\beta)&=&\lim_{K\rightarrow\infty}\int_{-\alpha}^{\infty}\prod_{k}d\lambda_{k}
    \frac{1}{(2\pi)^{K}}\int_{-\infty}^{\infty}\prod_{k}d\omega_{k}\nonumber\\
    &\times&\sum_{\vx}\exp{\Big(j\sum_{k}{\omega_{k}\lambda_{k}}\Big)}
    \cdot\mathbb{E},
\EEA where $j\triangleq\sqrt{-1}$ and \BE\label{eq_expectation}
    \mathbb{E}\triangleq\Bigg<\exp{\Big(-j\sum_{i\neq k}\frac{1}{K}\sum_{\mu=1}^{N}s_{k}^{\mu}s_{i}^{\mu}x_{i}x_{k}\omega_{k}\Big)}\Bigg>_{\rho}.\nonumber \EE
The expectation $\mathbb{E}$ can be also written as \BEA\label{eq_expectation11}
    \mathbb{E}&=&\exp{(j\alpha\sum_{k}\omega_{k})}\nonumber\\&\times&\Big<\exp{\big(-\frac{j}{K}\sum_{\mu}(\sum_{k}s_{k}^{\mu}x_{k}\omega_{k})(\sum_{k}s_{k}^{\mu}x_{k})\big)}\Big>_{\rho}.
\EEA
    Using a transformation~\cite[eq. (2.14)]{BibDB:BruceEtAl}, the expectation becomes
\BEA\label{eq_expectation2}
    \mathbb{E}&=&\exp{(j\alpha\sum_{k}\omega_{k})}\int_{-\infty}^{\infty}\prod_{\mu}\frac{da_{\mu}}{(2\pi/K)^{1/2}}\nonumber\\
    &\times&\int_{-\infty}^{\infty}\prod_{\mu}\frac{db_{\mu}}{(2\pi/K)^{1/2}}\exp{\Big(j\frac{K}{2}\sum_{\mu}(a_{\mu}^{2}-b_{\mu}^{2})\Big)}\nonumber\\
    &\times&\exp{\Big(\sum_{k,\mu}\log\big(\cos(c_{k,\mu})\big)\Big)},
\EEA where $c_{k,\mu}\triangleq\frac{1}{\sqrt{2}}\big(\omega_{k}(a_{\mu}+b_{\mu})+(a_{\mu}-b_{\mu})\big)$. Since
$\sum_{k}s_{k}^{\mu}x_{k}$ in~(\ref{eq_expectation11}) is $\mathcal{O}(\sqrt{K})$ for an overwhelming majority
of codewords, for the expectation $\mathbb{E}$ to be finite, $a_{\mu}$ and $b_{\mu}$ must be
$\mathcal{O}(1/\sqrt{K})$. Hence, expanding the $\log\big(\cos(\cdot)\big)$ term in
exponent~(\ref{eq_expectation2}) and neglecting terms of order $1/K$ and higher, we get
\BEA\label{eq_expectation3}
    \mathbb{E}&=&\exp{(j\alpha\sum_{k}\omega_{k})}\int_{-\infty}^{\infty}\prod_{\mu}\frac{da_{\mu}}{(2\pi/K)^{1/2}}
    \nonumber\\&\times&\int_{-\infty}^{\infty}\prod_{\mu}\frac{db_{\mu}}{(2\pi/K)^{1/2}}\exp{\Big(j\frac{K}{2}\sum_{\mu}(a_{\mu}^{2}-b_{\mu}^{2})\Big)}\nonumber\\
    &\times&\exp{\Big(-\frac{1}{4}\sum_{k,\mu}\hat{c}_{k,\mu}\Big)}, \EEA
where
$\hat{c}_{k,\mu}\triangleq\big(\omega_{k}^{2}(a_{\mu}+b_{\mu})^{2}+2\omega_{k}(a_{\mu}^{2}-b_{\mu}^{2})+(a_{\mu}-b_{\mu})^{2}\big)$.
Now, the multi-dimensional integral~(\ref{eq_expectation3}) is solved using the following mathematical recipe:
New variables are introduced\BEA
    a\triangleq\frac{1}{2\alpha}\sum_{\mu}(a_{\mu}+b_{\mu})^{2},\quad
    b\triangleq\frac{j}{2\alpha}\sum_{\mu}(a_{\mu}^{2}-b_{\mu}^{2})+1\label{eq_a}.
\EEA Equations~(\ref{eq_a}) can be reformulated via the integral representation of a delta function using the
corresponding angular frequencies $A$ and $B$, respectively, \BEA
    \int_{-\infty}^{\infty}\frac{da\ dA}{2\pi/K\alpha}\exp{\big(jKA(\alpha a-\sum_{\mu}\frac{(a_{\mu}+b_{\mu})^{2}}{2})\big)}=1,\nonumber\\
    \int_{-\infty}^{\infty}\frac{db\ dB}{2\pi/K\alpha}\exp{\big(jKB(\alpha b-j\sum_{\mu}\frac{(a_{\mu}^{2}-b_{\mu}^{2})}{2}-\alpha)\big)}=1\label{eq_bF}.\nonumber
\EEA Substituting these (unity) integrals into the expectation expression~(\ref{eq_expectation3}) and rewriting
it using $a$ and $b$, the integrations over $a_{\mu}$ and $b_{\mu}$ are decoupled and can be performed easily.
Next, for the asymptotics $K\rightarrow\infty$, the integration over the frequencies $A$ and $B$ can be
performed algebraically by the saddle-point method~\cite{BibDB:BookEllis}. According to this method, the main
contribution to the integral comes from values of $A$ and $B$ in the vicinity of the maximum of the exponent's
argument. Finally, the $\mathbb{E}$ term boils down to \BEA\label{eq_expectation_final}
    \mathbb{E}&=&\int_{-\infty}^{\infty}\frac{da\
    db}{4\pi/K\alpha}\exp{\big(K\alpha(b-\frac{1}{2}+\frac{(1-b)^{2}}{2a}+\frac{1}{2}\log{a})\big)}\nonumber\\
    &\times&\exp{\big(-\frac{1}{2}\alpha a\sum_{k}\omega_{k}^{2}+j\alpha b\sum_{k}\omega_{k}\big)}.
\EEA Substituting the expectation term~(\ref{eq_expectation_final}) back in~(\ref{eq_number3}), the integrand in
the latter becomes independent of $\vx$, therefore the $\sum_{\vx}$ can be substituted by multiplying with the
scalar $2^K$, and the resulting $\omega$ dependent integrand is a Gaussian function. Thus performing Gaussian
integration and exploiting the symmetry in the $K$-dimensional space, we get \BEA\label{eq_number4}
    &&\mathcal{N}(\beta)=\lim_{K\rightarrow\infty}\frac{1}{\pi^{K}}
    \int_{-\infty}^{\infty}\frac{da\
    db}{4\pi/K\alpha}\nonumber\\&\times&\exp{\Big(K\alpha\big(b-\frac{1}{2}+\frac{(1-b)^{2}}{2a}+\frac{1}{2}\log{a}\big)\Big)}\\
    &\times&\exp{\bigg(K\log{\Big(\sqrt{\frac{2\pi}{\alpha a}}\int_{-\alpha}^{\infty}d\lambda\exp{\big(-\frac{(\alpha b+\lambda)^{2}}{2\alpha
    a}\big)}\Big)}\bigg)}.\nonumber
\EEA Using the rescaling $(\alpha b+\lambda)/\sqrt{\alpha a}\rightarrow \lambda$, the
integral~(\ref{eq_number4}) becomes \BE\label{eq_integral}
    \mathcal{N}(\beta)=\lim_{K\rightarrow\infty}\int_{-\infty}^{\infty}\frac{da\
    db}{4\pi/K\alpha}\exp{\big(Kg(a,b,\beta)\big)},
\EE where the function $g(a,b,\beta)$ is defined by \BEA
    g(a,b,\beta)&\triangleq&\frac{1}{\beta}\big(b-\frac{1}{2}+\frac{(1-b)^{2}}{2a}+\frac{1}{2}\log{a}\big)\nonumber\\&+&\log\big(2Q(t)\big).\nonumber
\EEA The definitions of the auxiliary variable $t\triangleq\sqrt{\alpha}(b-1)/\sqrt{a}$ and the error function
$Q(x)\triangleq1/\sqrt{2\pi}\int_{x}^{\infty}dy\exp{(-y^{2}/2)}$ are used. Again, for $K\rightarrow\infty$, the
double integral in~(\ref{eq_integral}) can be evaluated by the saddle-point method. Hence, we find
\BE\label{eq_finalNumber}
    \mathcal{N}(\beta)\propto\lim_{K\rightarrow\infty}\exp{\big(Kg(a^{\ast},b^{\ast},\beta)\big)},
\EE where $a^{\ast}$ and $b^{\ast}$ are found by the saddle-point conditions, which yield the following
equations \BEA
    \pd{g(a,b,\beta)}{a}&=&\beta^{-1}\big(\frac{(1-b)^{2}}{a}-1\big)+t\frac{Q'(t)}{Q(t)}=0,\nonumber\\
    \pd{g(a,b,\beta)}{b}&=&\beta^{-1}\big(1-\frac{1-b}{a}\big)+\frac{1}{\sqrt{a\beta}}\frac{Q'(t)}{Q(t)}=0.\nonumber
\EEA The operator $Q'$ denotes a derivative of $Q$ w.r.t. its argument. One then finds that this set of
equations is satisfied by $b^{\ast}=0$ and \BE
    a^{\ast}=\beta^{-1}/\big({\beta^{-1}+\frac{1}{\sqrt{a^{\ast}\beta}}\frac{Q'(t^{\ast})}{Q(t^{\ast})}}\big),\label{eq_fix_a}
\EE where $t^{\ast}\triangleq-1/{\sqrt{a^{\ast}\beta}}$. This saddle-point condition's fixed-point $a^{\ast}$
can be found iteratively, and it always converges in the examined model~\cite{BibDB:Singh}.

Finally, substituting~(\ref{eq_finalNumber}) the asymptotic capacity, in nat per symbol per user, is now easily
obtained \BEA\label{eq_C}
    C_{\infty}(\beta)&=&g(a^{\ast},b^{\ast},\beta)=\log\big(2Q(t^{\ast})\big)\nonumber\\&+&\frac{1}{\beta}\big(b^{\ast}-\frac{1}{2}+\frac{(1-b^{\ast})^{2}}{2a^{\ast}}+\frac{1}{2}\log{a^{\ast}}\big),
\EEA which forms our pivotal result. In section~\ref{sec_results} we further discuss the theoretical results and
compare them with computer simulations of the complexity-constrained CDMA channel.

\section{Results}\label{sec_results}
\begin{figure}
\centering
\includegraphics[width=3in]{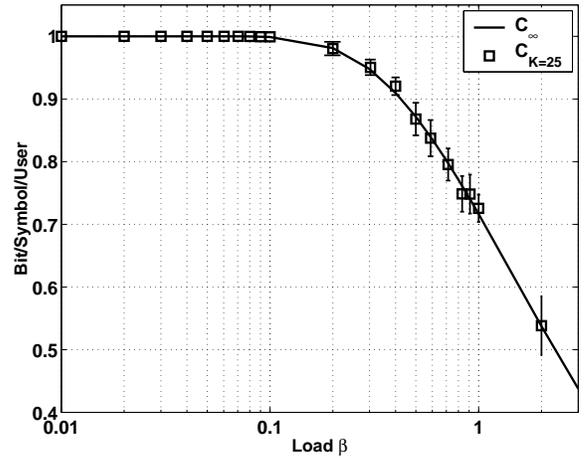}
\caption{Asymptotic capacity $C_{\infty}$ (solid line), in terms of bit/symbol/user, as a function of load
$\beta$. Also drawn is the finite-size simulation-averaged capacity $C_{K}$ for $K=25$ (empty squares). Vertical
bars stand for standard deviation in simulation results.} \label{fig_capacity}
\end{figure}

Fig.~\ref{fig_capacity} displays the asymptotic capacity $C_{\infty}$~(\ref{eq_C}), obtained by solving
iteratively the saddle-point condition~(\ref{eq_fix_a}), as a function of the load $\beta$. Interestingly, for
small $\beta\lesssim0.1$ values the trivial 1 bit upper bound (of an optimal receiver, \textsl{i.e.} matrix
inversion) is practically achieved by this simple hard decision operation. Nevertheless, even for higher
non-trivial system load such a complexity-constrained CDMA setting still yields substantial achievable
information rates. Note, in passing, that for heavily overloaded system (\textsl{i.e.} $\beta\rightarrow\infty$)
the capacity curve decay coincides with Hopfield model's capacity (see~\cite[eq.~(12)]{BibDB:Gardner} for an
analytical approximation of this capacity decay to zero.)

In order to validate the analytically derived asymptotic capacity $C_{\infty}(\beta)$, we evaluated the capacity
$C_{K}(\beta)$ of a CDMA downlink channel with large, yet finite number of users $K$, using exhaustive search
simulations. The number of successful binary codewords, maintaining the channel constraints, was obtained by
examining all $2^K$ possible codewords. The average logarithm of the counted number, normalized by the number of
users $K$, gives the capacity $C_{K}$. Fig.~\ref{fig_capacity} presents the capacity obtained by simulations for
$K=25$. As can be seen, the empirical capacity for finite $K$ deviates only slightly from the analytically
obtained asymptotic capacity. These results substantiate the analysis of the complexity-constrained CDMA
channel.

\section{Concluding Remarks}\label{sec_conclusion}
We evaluated the asymptotic capacity of a CDMA downlink channel model requiring only minimal signal processing
at the receiver, thus suitable for interference-limited systems with low-complexity constrained mobile
equipment. Interestingly, we found a range of non-trivial achievable rates. According to these findings, at a
given channel use a fraction of the users, equal to $C_{\infty}$ (in bit), can receive its designated
information with rate $1$, while the transmissions to the rest of the users ensure reliable communication.
Determining these redundant transmissions in a diagrammatic manner (rather than via brute-force enumeration,
which becomes infeasible for large $K$) remains an interesting open research question.

\section*{Acknowledgment}
The authors are grateful to Shlomo Shamai (Shitz) for useful discussions and the anonymous reviewers for
valuable comments. O.S. wishes to thank Noam Shental for constructive comments on the manuscript.

\end{document}